\documentclass[conference]{IEEEtran}
\IEEEoverridecommandlockouts
\usepackage{amsmath,amssymb,amsfonts}
\usepackage{algorithmic}
\usepackage{graphicx}
\usepackage{textcomp}
\usepackage{xcolor}
\usepackage{stfloats}
\usepackage{float}
\newtheorem{dfn}{Definition}
\usepackage{times}  
\usepackage{helvet} 
\usepackage{courier}  
\usepackage[hyphens]{url}  
\usepackage{subfigure}
\usepackage{hyperref}
\usepackage{graphicx} 
\usepackage{bm}
\usepackage{xcolor}
\usepackage{amsfonts}
\usepackage{multirow}
\usepackage{tikz}
\usetikzlibrary{positioning, calc}
\newcommand{\U}{\mathcal{V}}
\newcommand{\PM}{{\it PM}}
\newcommand{\ca}[1]{\mathcal{#1}}
\usepackage{multirow}
\def\BibTeX{{\rm B\kern-.05em{\sc i\kern-.025em b}\kern-.08em
    T\kern-.1667em\lower.7ex\hbox{E}\kern-.125emX}}

\begin{document}

\title{From \#Jobsearch to \#Mask: Improving COVID-19 Cascade Prediction with Spillover Effects
}

\author{
    \IEEEauthorblockN{
        Ninghan Chen\IEEEauthorrefmark{1}, 
        Xihui Chen\IEEEauthorrefmark{2}, 
        Zhiqiang Zhong\IEEEauthorrefmark{1}, and
        Jun Pang\IEEEauthorrefmark{1}\IEEEauthorrefmark{2} \\[5pt]
        }
    \IEEEauthorblockA{
        \begin{tabular}{cc}
            \begin{tabular}{@{}c@{}}
                \IEEEauthorrefmark{1}
                    Faculty of Sciences, Technology and Medicine\\
                    University of Luxembourg\\
                    Esch-sur-Alzette, Luxembourg\\
            \end{tabular} & \begin{tabular}{@{}c@{}}
                \IEEEauthorrefmark{2}
                    Interdisciplinary Centre for Security, Reliability and Trust\\
                    University of Luxembourg\\
                    Esch-sur-Alzette, Luxembourg\\
            \end{tabular}
        \end{tabular}
    }
}
\maketitle

\begin{abstract}
An information outbreak occurs on social media along with the COVID-19 
pandemic and leads to \emph{infodemic}. 
Predicting the popularity of online content, known as \emph{cascade prediction}, 
allows for not only catching in advance hot information that deserves attention, but also 
identifying false information that will widely spread and require quick response
to mitigate its impact.  Among the various
information diffusion patterns leveraged in previous works,  
the \emph{spillover effect} of the information exposed to users on  
their decision to participate in diffusing certain information is still not studied.
In this paper, we focus on the diffusion of information related to COVID-19 preventive measures. 
Through our collected Twitter dataset, we validated the existence of this 
spillover effect.
Building on the finding, we proposed extensions to three cascade
prediction methods based on Graph Neural  Networks (GNNs). Experiments conducted on our dataset 
demonstrated that the use of the identified spillover effect significantly improves 
the state-of-the-art GNNs methods in predicting the popularity of not only preventive
measure messages, but also other COVID-19 related messages.
\end{abstract}

\section{Introduction}
\label{sec:introdution}

The outbreak of the COVID-19 pandemic leads to an outbreak of information in 
major online social networks (OSNs), including Twitter, Facebook, Instagram, 
and YouTube~\cite{cinelli2020covid}, which is called \emph{infodemic}. On one hand,  
due to physical isolation and social distancing, people spent much more time on 
OSNs, engaging in expressing opinions, catching 
up-to-the-minute development of the pandemic and even looking for 
medical support and knowledge to ease mental depression and 
seek psychological comfort.  
This new change in information perception makes OSNs 
become an essential communication channel for healthcare departments and 
medical staff to disseminate official policies and professional advice  
about effective measures to prevent the spread of COVID-19 virus, e.g.,  
wearing masks, vaccination and social distancing. 
Misinformation and false news also take advantage of 
social media to spread with unprecedented speed and volume.
Large-scale dissemination of misinformation significantly misleads people 
and causes public panic. As a result, this information explosion on social media 
hinders effective pandemic response and increases public confusion about who 
and what preventive measures to trust~\cite{alang2020police}.
One widely accepted solution to combat infodemic is known as \emph{cascade prediction}. 
Its purpose is to learn the popularity of messages given its early adopters.
Accurate prediction can help people catch hot 
information that deserves attention and assist healthcare department identify misinformation 
that will require fast response to control the impact in advance. 

Research on cascade prediction has been sustained, with a large number of
prediction models developed. Earlier models rely on hand-crafted features extracted 
from demographic profiles of early 
adopters~\cite{cheng2014can,cui2013cascading,jenders2013analyzing} 
and the subgraphs composed of early adopters and their relationships~\cite{mooney1997monte}. 
The recent advances of representation learning techniques lead to end-to-end 
representation-based prediction models~\cite{bourigault2016representation,gao2017novel}. 
Particularly, the application of graph neural networks (GNN) allows to 
simulate cascading effects over social networks, and further improves the performance of 
cascade prediction~\cite{mooney1997monte}. 
In spite of the various diffusion patterns exploited, the works mentioned so far have not
considered the \emph{spillover effect} of a user's exposed information over 
social media on his/her behaviour of forwarding a message and becoming part 
of its diffusion, which we call \emph{info-exposure spillover effect} for short. 
We say a user is \emph{exposed} to a message if 
the user posted the message or perceives it from his friends on social media. 
Here, we adopt the definition of behaviour spillover effect which 
intuitively means \emph{``the observable and 
causal effect that a change in one behaviour has on a different, subsequent 
behaviour''}~\cite{GW19}. 
For example, tweets about unemployment and job-searching may make  
a user who read them perceive the severity of the pandemic and 
thus more likely retweet tweets about preventive measures like stay-at-home. 

We hypothesise the existence of this info-exposure spillover effect according to the previous 
studies related the COVID-19 pandemic.   
Park et al.~\cite{park2020conversations} demonstrate that information with 
medically oriented thematic framework has a wider spillover effect on 
COVID-19 issues in a Twitter context. 
Racist information can have a spillover effect on the mistrust of medical 
system~\cite{alang2020police} and lead to a lack of trust in the information 
released by these systems. 
In this paper, we focus on the messages related to COVID-19 preventive measures
considering their importance in the combat against the pandemic. 
We collected a dataset from Twitter and 
successfully validated the existence of the info-exposure spillover effect of 
users' exposed messages on their decision to retweet messages related to preventive measures. 
This allows us to extend existing state-of-the-art cascade prediction
models relying on GNNs. 

According to our evaluation on our dataset, our 
extended models can increase the cascade prediction performance up to 23\% in COVID-19 
messages related to preventive measures. Meanwhile, we observed that the use of 
info-exposure spillover effect can also increase the accuracy in predicting the popularity 
of other COVID-19 related messages.

\section{Related Work}
\label{sec:Related work}

Cascade prediction has become attractive after studies shed
light on some key properties of information cascades that can be 
predicted~\cite{cheng2014can,yu2015micro}. In general, the cascade prediction methods
can be divided into two classes: macro-level prediction and micro-level prediction. 
Micro-level prediction aims to predict users who will be activated during the information 
diffusion, while macro-level cascade prediction directly calculates the final size of 
targeted cascades. 

The idea of most micro-level methods are based on the Independent Cascade model 
(IC)~\cite{kempe2003maximizing}, which calculates the probability of influence between 
every pair of users~\cite{goldenberg2001talk,gomez2012inferring}. 
These methods rely on a number of assumptions that overly simplify the real situation such 
as the complete observation of diffusion processes~\cite{convexity2010Leskovec}. 
Although Deepinf~\cite{qiu2018deepinf} uses an end-to-end deep learning method to 
overcome such assumptions, micro-level methods generally do not perform 
well in predicting cascade future size as they require simulating the entire diffusion process.
In this paper, as our target is popularity prediction, we opt for macro-level methods.

Macro-level prediction methods can be divided into three categories as a result of 
technological evolution, i.e., statistical prediction model, 
machine learning-based methods and deep learning-based methods. 
The development of macro-level prediction starts with statistical models such as 
SEISMIC~\cite{zhao2015seismic} and Weibull~\cite{yu2015micro}. Then, 
the advancements of machine learning lead to methods using manually designed features 
extracted from text content, temporal and demographic information, and network 
structure~\cite{yu2015micro,cheng2014can,cui2013cascading}. 
Deep learning-based methods overcome the deficiency of machine learning-based methods 
of constructing manual features and capture effective features automatically. 
DeepCas~\cite{li2017deepcas} and DeepHawkes~\cite{cao2017deephawkes} 
use Recurrent Neural Networks (RNNs) to capture cascading sequences in place of manually 
designed features. However, RNNs are limited in capturing structural information. 
This limitation is addressed by graph neural networks 
(GNNs)~\cite{scarselli2008graph}. Intuitively,  
GNNs update the representation of each node by recursively aggregating the 
representations of its neighbours. In this way, the iterated node representation 
summarises both structural and representation information in neighbourhoods.
CasCN~\cite{chen2019information} utilises a dynamic Graph Convolutional Network (GCN) 
to learn the structural information of the cascade. 
CoupledGNN~\cite{cao2020popularity} (CGNN) effectively addresses cascade prediction 
with two GNNs, capturing the cascading effect which indicates that the activation of 
one user will successively trigger its neighbours. 

Although deep learning-based methods have achieved relatively good results in 
cascade prediction, little research has been conducted to incorporate textual content 
into cascade prediction.  
Textual content, an important part of social media, may contain information 
that are related to the diffusion of messages. 
information a node has received in the past and its activate status. 
Thus, we narrow the focus in this article to macro-level cascade prediction by 
extending the existing models to explore online textual content. 

\section{Preliminaries }
\label{sec:preliminaries}

\subsection{Problem definition}
\label{sec:Problem Definition}
In this section, we will give the formal definition of the popularity prediction problem 
studied in this paper which takes into account both social networks and online textual 
contents. 

When a message $m$ is firstly posted by a user, it will be perceived by the user's followers  
who will adopt the message and relay the message. 
This cascading process will continue on the social network until no further sharing occurs.   
We denote the observed diffusion cascade of $m$ at in the time window $T$ by 
$C_m^T= \{u_1, u_2, \ldots, u_{n_T^m}\}$, i.e., the set of users who adopted $m$ 
in time window $T$. Note that $n_T^m$ is the number of the adopters of $m$ in time window $T$. 
We use graph $\ca{G}=(\ca{V},\ca{E})$ to denote the social network where 
$\ca{V}$ is the set of nodes representing users and $\ca{E}\subset\ca{V}\times \ca{V}$ 
is the set of edges indicating the relationships between users. 
Compared to the previous works on cascade prediction, we 
take into account the online textual messages posted by users. 
Specifically, for a user $v\in\ca{V}$,
given a time period, we use $\ca{M}_v$ to denote the messages posted by user $v$
and $\ca{M}$ to denote the set of all messages, i.e., $\ca{M}=\cup_{v\in\ca{V}}\ca{M}_v$.  

\smallskip
\noindent\textbf{Online textual content-aware cascade prediction.}
Given the cascade of message $m$ in time window $T$ (i.e., $C_m^T$), 
social network $\ca{G}=(\ca{V},\ca{E})$ 
and the messages posted by users in $\ca{V}$, i.e., 
$\forall_{v\in \ca{V}} \mathcal{M}_v$, 
the purpose of the problem is to predict the final popularity of $m$ at 
time $\infty$, i.e., $n_{\infty}^m$.
\smallskip

As mentioned previously, we focus on the diffusion of the messages related to COVID-19 
preventive measures in this paper and the user generated messages are also related 
to the COVID-19 pandemic. We will make use of the info-exposure spillover effects of
users' exposed information on their decision on relaying preventive measure messages 
to solve the cascade prediction problem. 
The result of this paper may be applicable to other types of 
information if similar spillover effect also exists. 

\subsection{General Framework of GNNs}
The purpose of Graph neural networks (GNN) is to calculate a representation of a graph. 
Compared to graph embedding works such as node2vec~\cite{grover2016node2vec} and DeepWalk~\cite{perozzi2014deepwalk}, 
one advantage of GNN is that it allows to integrate node attributes into the 
learning process.  GNN is implemented with multiple layers. At each layer, 
a node's embedding is updated by combining the representation of their neighbours calculated
in the previous layer. Intuitively, a $k$-layer GNN calculates a representation for 
each node by combining the attributes of the nodes within $k$ hops. We adopt the 
formal definition in~\cite{scarselli2008graph} and give the general definition of the $\ell$-th layer for 
a node $v\in\ca{V}$ as  follows:
\begin{displaymath}
\begin{array}{ll}
a_v^{\ell} &= {\sf Aggregate}(\{h_u^{\ell}: u\in \mathcal{N}(v)\})\\
h_v^{\ell+1} & = {\sf Combine}(h_v^{\ell}, a_v^\ell) 
\end{array}
\end{displaymath}
where $h_v^{\ell}$ is the representation vector of node $v$ at the $\ell$-th layer and 
$\ca{N}(v)$ denotes the neighbours of node $v$. 
Function ${\sf Aggregate}$ and ${\sf Combine}$ are instantiated in many variants of GNN
so as to capture different features of nodes' neighbourhoods.  
With the representation vector of every node at the $k$-th layer, then the representation 
of the graph $\mathcal{G}$ can thus be calculated by a function as follows:
\[
h_{\mathcal{G}} = {\sf Readout}(\{h_v^{k}: v\in \U\}).
\]
The {\sf Readout} function can be simply implemented as the mean of nodes' vectors or other 
complex pooling functions.
\section{Data Collection and Pre-processing }
\label{sec:data}
Twitter, one of the most prominent online social media platform, has been used extensively 
during the COVID-19 pandemic. We chose the Greater Region (GR)\footnote{The Greater Region 
of Luxembourg is composed of the Grand Duchy of Luxembourg, Wallonia, Saarland, 
Lorraine, Rhineland-Palatinate and the German-speaking community of Belgium.}, 
a region with a popularity of high mobility, as the targeted area in this paper. 
This section presents how we build the dataset, construct the cascades and build the 
social graph for our analysis and experiments. 

\subsection{Data collection}
In our dataset, we collect two types of data:   
i) the COVID-19 related tweets posted or re-tweeted by GR users; 
ii) the social networks of GR users recording their following relationships. 
In what follows, we elaborate the three steps we followed to gather these data.

\smallskip\noindent
\textbf{Step 1. Tweet collection.} 
At this step, we collect a set of seed users in GR who actively
participate in COVID-19 discussions and the tweets they originally posted or retweeted. 
Instead of searching by keywords, we refer to a publicly available dataset 
which contains the IDs of COVID-19 related tweets~\cite{COVID-19Dataset}. 
We extract the tweet IDs posted between  
January 22, 2020) and July 18, 2020. 
This period covers the first wave of the pandemic.
Through these IDS, we downloaded the corresponding tweet. 
Due to the ambiguity of locations of tweet posters, 
we use the geocoding APIs, Geopy 
and  ArcGis Geocoding
to regularise locations associated with tweets. 
For example, a user input location \emph{Moselle} 
is transformed to a preciser and machine-parsable location: 
\emph{Mosselle, Lorraine, France}.
Based on the regularised locations, we filter the downloaded tweets and 
remove those posted by users out of GR. 
In total, we obtain 144,961 tweets from 8,872 GR users. 

\smallskip\noindent
\textbf{Step 2. Social network construction.}
We construct the social network of a large number of GR users at this step. 
We use an iterative approach to gradually enrich the social network.
For each seed user, we
obtain his/her followers and only retain those who have a mutual following
relation with the seed user, 
because such users are more likely to reside in GR. 
We then download new users' locations from their profile data 
and only add users from GR to the social network.
We also add edges if users in the network have following relation 
with the newly added users. 
After the first round, we continue going through the newly added users 
by adding their mutually followed friends that do not exist in the 
current social network. This process will continue until no new users can be 
added. 
In our collection, it takes 5 iterations before termination.
We take the largest weakly connected component of the social network.
After this step,  we  collected  a  total  of  12,256,152  users
and 21,203,130 following relationships.
Since the majority of users in the network are relatively inactive, we 
construct a subgraph by removing all users who post or retweet less than 2 tweets. 
Note that we keep some of such inactive users when the remaining network are no longer 
connected after the removal of these users. 
In the end, we obtain a social network with 21,339 users and 214,962 edges.

\subsection{Cascade construction and experiment data selection} 
We construct cascades from our tweet dataset and the social network built previously 
based on the definition in section~\ref{sec:Problem Definition}. 
A total of 60,035 cascades are built and we remove cascades with less than 3 users, 
following the existing works~\cite{li2017deepcas,cao2020popularity}.  
Eventually, 82.38\% of cascades are removed and we ended up with 10,579 cascades.
The average size of these cascades is 4.78. We use $\ca{C}$ to denote the set of 
all selected cascades. From $\ca{C}$, we construct the set of cascades corresponding 
to messages related to preventive measures, denoted by $\ca{C}_\PM$, based on the 
keywords listed in Table~\ref{table:Keywords}.

\begin{table*}[hbt]
\caption{Keywords for the topic categories}
\label{table:Keywords}
\centering
\begin{tabular}{|l|l|l|l|l|}
\hline
 & \multicolumn{4}{l|}{\textbf{Keywords}} \\ \hline\hline
Preventive measure & \multicolumn{4}{l|}{\begin{tabular}[c]{@{}l@{}}stayathome, mask, masque, maske, washhand, wash hand, social distancing, socialdistancing, staysafe,  lockdown\end{tabular}} \\ \hline
 
Unemployment & \multicolumn{4}{l|}{\begin{tabular}[c]{@{}l@{}}job, jobsearching, jobsearch, unemployment, employment, career, resume, recruitment, recession, economy, economic emploi,\\ stelle, employ,  arbeitslos,  chômeurs\end{tabular}} \\ \hline
Panic buying & \multicolumn{4}{l|}{\begin{tabular}[c]{@{}l@{}}panicbuying, panicshopping, panicbuyers, toiletpaper, handsanitizer, coronashopping\end{tabular}} \\ \hline
School closures & \multicolumn{4}{l|}{\begin{tabular}[c]{@{}l@{}}schoolclos,  closenypublicschools, closenycschools, suny, cuny, homeschool, noschool,  closetheschools, shutdownschools\end{tabular}} \\ \hline
\end{tabular}
\end{table*}

\section{Spillover effects in COVID-19 preventive measure information diffusion}
\label{data_analysis}

In this section, we will validate our hypothesis that the information exposed to a user
has a spillover effect on his/her behaviour of adopting a message related to COVID-19 preventive
measures. We first briefly describe the method we use to measure the hypothesised info-exposure 
spillover effect
and then give the detailed experimental analysis designed to validate its existence in 
the diffusion of  COVID-19 preventive measure messages.  

\subsection{Measuring info-exposure spillover effect}

We design our experimental framework based on the experimental investigation method 
for spillover effect validation. Intuitively, the idea is to investigate whether users 
exposed to different information will behave differently in retweeting a message related to 
preventive measures. In other words, 
we will check whether certain type of exposed information will change the likelihood of 
users with regard to retweeting a preventive measure message. 

We divide the set of users into groups according to the information they are exposed to.
Each group is composed of users who are exposed to a certain composition of information. 
One of this groups is set as the control group. The selection of the control group 
depends on the purpose of the experiment. Then the proportion of users in each group 
retweeting preventive measure messages will be used to estimate the likelihood 
of adopting preventive measure messages. By comparing the measurements with the control group, 
we can then quantitatively evaluate the gratitude of the spillover effect of the information 
exposed to this user group on adopting preventive measure messages, which we call 
\emph{spillover elasticity}. 

Formally, the nodes in social graph $\ca{G}$ will be divided into $n$ groups, i.e., 
$\mathcal{D} = \{\U_1, \ldots, \U_n\}$ where $\cup_{\U'\in\mathcal{D}}\U'=\U$. 
Let $\U_c\in\mathcal{D}$ be the selected control group.
For each user group $\U_i$, we will find the users who re-tweet preventive measure 
messages in $\mathcal{M}_\PM$, and construct the set of users $\U_i^\PM$. Then the activation 
likelihood for users in $\U_i$ is calculated as 
$$\alpha_{\U_i}=\frac{\mid\U_i^\PM\mid}{\mid\U_i\mid}.$$ 
With these notations, we can define spillover elasticity. 
\begin{dfn}[Spillover elasticity]
The \emph{elasticity} of the info-exposure spillover effect of 
group $\U_i$ in a division $\mathcal{D}$ of user set $\U$ 
is calculated as 
\[
\varepsilon_{\U_i}^\mathcal{D} = \frac{\alpha_{\U_i} - \alpha_{\U_c}}{\alpha_{\U_c}}.
\]
\end{dfn}
Positive elasticity indicates the exposure to the information of $\U_i$  
increases the likelihood of adopting a preventive measure message while negative elasticity 
indicates the opposite. 

\subsection{Experimental validation of info-exposure spillover effect}

We start to verify that being exposed to 
certain information will affect users' behaviour of adopting and re-tweeting 
COVID-19 preventive measure messages.
In order to conduct our experimental analysis, 
we need to first distinguish the types of COVID-19 related information. 
Previous studies~\cite{mamun2020covid,shanthakumar2020understanding} 
classified COVID-19 related information into several topics. 
Among these topics, we select three that are widely discussed in our dataset, 
i.e., \emph{unemployment}, 
\emph{panic buying} and \emph{school closures}, and extract corresponding tweets with 
the keywords listed in Table~\ref{table:Keywords}.

We conduct our analysis from two perspectives. We first evaluate the influence of 
messages of a single topic on the behaviour of adopting a preventive measure 
message. Second, we investigate the influences of different 
compositions of topics of messages.

\smallskip\noindent
\textbf{Spillover effect of information of single topic.}
We build three divisions of the users in order to evaluate 
the spillover effect of each topic, i.e., $\mathcal{D}_{U}$, 
$\mathcal{D}_{P}$ and $\mathcal{D}_{S}$ for unemployment, panic buying and 
school closure, specifically. Each division has only two groups. 
One group consists of users that have been exposed to messages of the corresponding 
topic while the other group is composed of users that have not been exposed.
We will take the group unexposed to the topic of messages as the control group.
In table~\ref{table:Percentage}, we summarise the results about the number of users 
exposed and unexposed in each division and the activation likelihood as 
well as the final elasticity.  

We have two observations from this table. First, the exposure to each type of messages will 
increase the likelihood of users to re-tweet a preventive measure message. On average, the 
activation likelihood equals to 0.56 for the exposed group while the unexposed group 
only has an activation likelihood of 0.26. The average elasticity is 1.27, which indicates that 
the activation likelihood doubles for the users exposed to the topics we selected on average. 
Second, the increase of activation likelihood for exposed users differs among the topics of 
exposed information. For instance, the exposure to information related to panic buying just 
leads to 25\% increase which is much smaller than the other two topics of messages. 

For the above analysis, we can conclude that i) exposure to certain topics of information will 
have a positive spillover effect on users' adopting preventive measure messages; 
ii) the scale of spillover effect differs according to the topics of exposed messages.

\begin{table}[ht]
\caption{Validation of info-expoure spillover effect of single topics.}
\label{table:Percentage}
\centering
\begin{tabular}{|l|r|r|r|r|r|}
\hline
\multirow{2}{*}{} & \multicolumn{2}{l|}{\textbf{Exposed}} & \multicolumn{2}{l|}{\textbf{Unexposed}} & \multicolumn{1}{l|}{\multirow{2}{*}{\textbf{Elasticity} $\varepsilon$}} \\ \cline{2-5}
 & \multicolumn{1}{l|}{\#user} & \multicolumn{1}{l|}{$\alpha$} & \multicolumn{1}{l|}{\#user} & \multicolumn{1}{l|}{$\alpha$} & \multicolumn{1}{l|}{} \\ \hline\hline
Unemployment & 4,238 & 0.67 & 17,101 & 0.25 & 1.69 \\ \hline
Panic buying & 6,119 & 0.39 & 15,220 & 0.31 &0.25 \\ \hline
School closures & 6,460 & 0.61 & 14,879 & 0.21 & 1.87 \\ \hline
\end{tabular}
\end{table}

\begin{figure}[ht]
\centering
\includegraphics[scale = 0.7]{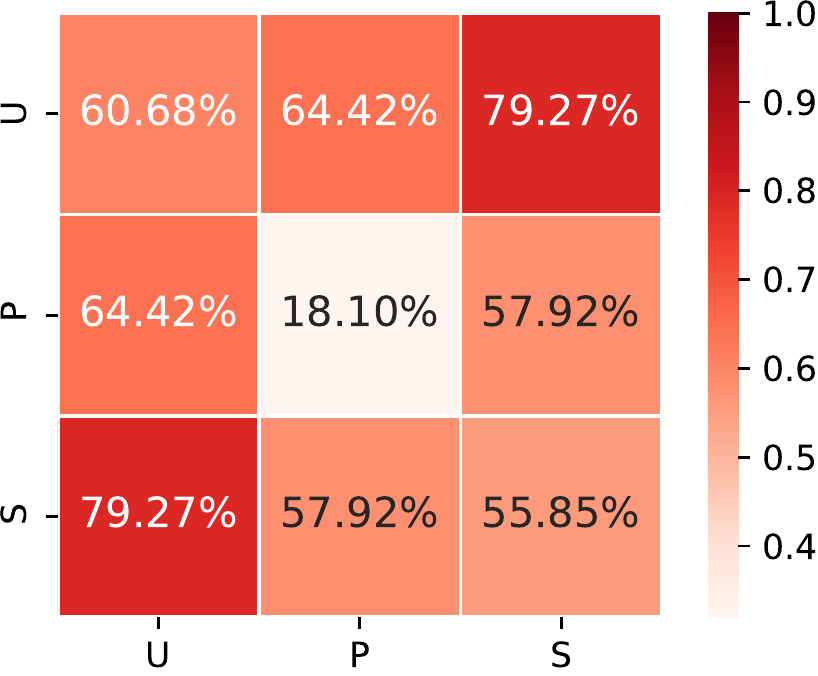}
\caption{Activation likelihood when exposed to compositions of topics.}
\label{fig:heatmap}
\end{figure}

\smallskip\noindent
\textbf{Spillover effect of information of compositions of topics.}
In the previous analysis, we focus on the spillover effect of single topics and ignore 
the changes when multiple topics of information are exposed to users simultaneously.
We construct of a division of the users $\mathcal{D}_{\it comp}$ with 8 groups, 
each of which corresponds to a possible composition of the three selected topics, 
i.e., $\{U\}, \{P\}, \{S\}, \{U, P\}, \{U, S\}, \{P,S\}, \{U,P,S\}, 
\emptyset$ where $U$, $P$ and $S$ are short for unemployment, panic buying and 
school closure, respectively. The user group exposed to none of the topics is 
selected as the control group.  Figure~\ref{fig:heatmap} shows the activation likelihood 
of user groups exposed to one or two selected topics of messages. 

We can see that exposure to more selected topics increases the likelihood 
of adopting a preventive measure message. The most significant increase occurs 
to the panic buying topic. Exposure to an additional topic will increase the activation 
likelihood by more than two times. When exposed to all the topics, the activation 
likelihood is increased to $0.81$. When exposed to none of the topics,  
the activation likelihood for the users drops below $0.10$.

From the analysis, we empirically validated that the information exposed to users 
indeed has a spillover effect on the behaviour of adopting a
preventive measure message. In other words, the likelihood of a user to 
re-tweet a preventive measure message will differ if they are exposed to 
different information. In the following, we will make use of this phenomenon 
to improve the accuracy of the prediction of message popularity. 
\section{Predicting popularity of COVID-19 preventive measure messages with spillover effect}
\label{sec:method}
In this section, we will make use of our findings of the spillover effect of users' 
exposed information on their decision of retweeting preventive measure messages to 
improve the accuracy of cascade prediction. 
Recall that the information exposed of users is composed of the 
messages posted by their friends and his/her own posts. 
We encode users' posted textual messages into representation vectors and 
then attach them to the attributes of the corresponding nodes in the social network.
Then we can make use of the GNN framework to summarise the messages posted
by their neighbours and even users that are not incident but within a certain
number of hops defined by the number of layers in GNNs. In the following, 
we start with describing the node attributes of the social graph and then proceed 
to extend GNN-based models to integrate info-exposure spillover effect. We also give 
the objective function to train the model parameters

\subsection{Preparing initial node attributes} 
The node attribute of node $v\in V$, i.e., $h_v^0$ is concatenated by three components: 
i) $s_v$, the activation status of the user 
in the given cascade $C$, ii) $\delta_v$, the representation vector of the messages posted by 
the user, and iii) $e_v$, the node embedding of the user's corresponding node in the network. 
Formally,  $h_v^0 = s_v\Vert \delta_v\Vert e_v$ 
where $\cdot\Vert\cdot$ is the concatenation operator. 
The user activation status $s_v$ is set to 
$1$ if $v\in C$ and $0$, otherwise. The node embedding captures the structural properties of 
the user's neighbourhood in the graph. Following existing 
studies~\cite{li2017deepcas,cao2020popularity}, we use DeepWalk without 
further fine-tuning to learn the structural embedding for each user. 
We will describe in detail the method to abstract the messages posted by the user 
into a representation vector.
RoBERTa~\cite{liu2019roberta} is a language pre-trained transformer to encode short texts in
multiple languages. 
In this paper, we use a widely used multilingual pre-trained RoBERTa variant: 
XLM-RoBERTa~\cite{conneau2019unsupervised}. 
For each $m\in\mathcal{M}$, we calculate its embedding with our trained XLM-RoBERTa model, and 
let $d_m$ be the corresponding embedding vector. For the messages posted by user $v$, we take 
the mean of the embedding vectors of all his posted messages as the final message embedding.
Formally, we have 
$\delta_v = \frac{1}{\mid \mathcal{M}_v\mid}\sum_{m\in\mathcal{M}_v}d_m$.

\subsection{Instantiating GNNs with info-exposure spillover effect}
%
We implement three variants of GNN to integrate the info-exposure spillover effect 
we identified in the previous section, i.e., Graph Convolutional Networks (GCN)~\cite{KipfW17},
Graph Attention Network~\cite{abs-1710-10903} and CoupledGNN~\cite{cao2020popularity}. 
GCN is a semi-supervised learning algorithm for graph representation and GAT is a variant of GCN 
which introduces the attention mechanism to distinguish the significance of neighbours. 
These two variants are not designed specifically for cascade prediction, but 
for the general purpose of summarising neighbourhoods
with a given depth. The calculated node representation can then be used for the 
downstream tasks such as link prediction and node classification. 
CoupledGNN~\cite{cao2020popularity} is a model developed for cascade prediction, and can stand 
for the state-of-the-art. 
It has overwhelming performance over existing models by considering the cascading 
effect of information diffusion on social network, i.e., the phenomenon that users are 
activated  due to the influence from their activated neighbours. 

By extending these models, our purpose is to illustrate the effectiveness of info-exposure 
spillover effect in improving further the performance of predicting the popularity of 
COVID-19 preventive measure messages. In addition, our extension can provide useful 
references for future cascade prediction models to integrate info-exposure spillover effect. 

\begin{table*}[th]
\caption{Brief description of selected GNN variants.\label{tab:gnnvariants}}
\centering 
\begin{tabular}{|l|l|l|}
\hline
\textbf{Model} & \textbf{\sf Aggregate(*)} & \textbf{\sf Combine(*)} \\\hline\hline
\textbf{GCN} &  
$\bm{a}_v^{\ell} = \frac{\sum_{u\in\mathcal{N}(v)\cup\{v\}}\bm{h^{\ell-1}_u}}{\mid\mathcal{N}(v)\cup\{v\}}$ & 
$\bm{h}_v^{\ell} ={\sf LeakyReLu} \left(\bm{W}^{\ell}\bm{a}_v^{\ell}\right)$ 
\\\hline
\textbf{GAT} & 
$\begin{array}{ll}
&\bm{a}_v^{\ell} = \sum_{u\in\mathcal{N}(v)\cup\{v\}}\bm{\beta_{uv}^{\ell}h^{\ell-1}_u} \\
& \\
&\bm{\beta^\ell_{uv}}  = \frac{\exp\left({\sf LeakyRelu} (\bm{\gamma}^T[\bm{Wh_u^{\ell-1}}\Vert \bm{Wh_v^{\ell-1}}])\right)}
{\sum_{u'\in\mathcal{N}(v)\cup\{v\}}\exp\left({\sf LeakyRelu} (\bm{\gamma}^T[\bm{Wh_{u'}^{\ell-1}}\Vert \bm{Wh_v^{\ell-1}}])\right)}
\end{array}$
& 
$\bm{h}_v^{\ell} ={\sf LeakyReLu} \left(\bm{W}^{\ell}\bm{a}_v^{\ell}\right)$ 
  \\\hline
\textbf{CoupledGNN} &  
$\begin{array}{ll}
&\bm{a}_v^{\ell} = \sum_{u\in\mathcal{N}(v)}{\sf InfluGate}\left(\bm{r_u^{\ell-1},r_v^{\ell-1}\right)s^{\ell-1}_u} +p_v \\
& \\
&{\sf influGate}\left(\bm{r_u^\ell}, \bm{r^\ell_v}\right) =
\bm{\beta^\ell}\left[\bm{W^\ell r_u^\ell}\Vert \bm{W^\ell r_v^\ell}\right]
\end{array}$
& $\bm{s^{\ell+1}_{v}}  =\begin{cases}
  1 & v\in C_m^T\\    
  \sigma(\bm{\mu_s^\ell s_v^\ell} + \bm{\mu_a^\ell a_v^\ell}) & v\not\in C_m^T    
\end{cases}  $
\\\hline 
\end{tabular}
\end{table*}

The definitions of the function {\sf Aggregate(*)} and {\sf Combine(*)} of 
GCN, GAT and CoupledGNN are briefly given in Table~\ref{tab:gnnvariants}.  
GAT and GCN share the same combination function.  
For GCN, we use the mean of the representation vectors of both the nodes and 
their one-hop neighbours as the aggregated value at each layer while 
GAT uses the weighted average. 

We describe CoupledGNN in more details due to its simulation of the cascading 
effect in information diffusion. For the full description, we refer the readers to 
the original paper~\cite{cao2020popularity}.
It deploys two GNNs. One GNN captures the 
activation statuses of users during the information diffusion at each layer, 
e.g., the activation status of user $v$ at the $\ell$-th layer $s_v^\ell$.
The other GNN aims to simulate the influence of users changing along with the
activation status and the influences of their neighbours, i.e., $r_u^\ell$.  
A neighbour $u$'s influence to user $v$ on becoming active in the next layer $\ell+1$ 
is calculated by the function ${\sf influGate}(\bm{r_u^\ell},\bm{r_v^\ell})$.
Then the aggregation function is the weighted average of all the neighbours' activation 
statuses with the default activation probability $p_v$ added. 
The combination function is based on the weighted average of its status of the previous 
layer and the aggregated representation. With the output activation status at 
the last layer (e.g., $k$), the popularity of the message diffused in $C^m_T$ is 
calculated as $$\Tilde{n}_\infty^m=\sum_{v\in\mathcal{V}} s_v.$$
In the following, we will describe in detail how we extend each selected model 
to capture the info-exposure spillover effect.

\medskip\noindent\textbf{SE-GCN \& SE-GAT.} 
We can interpret the output of the $k$-th layer of a $k$-layered GCN or GAT as 
the summary of the information exposed to every user. Then we use an activation function
to capture the info-exposure effect. 
Specifically, the function 
takes as input the output of the GCN or GAT and the representation of 
the message diffused in the given cascade and outputs the predicted final activation 
status of the nodes. Let $m$ be the message being diffused and recall that $d_m$
is the representation of $m$ calculated by the RoBERTa model.
Let $\Tilde{s}_v^\infty$ be the predicted activation status of node $v$.
Our activation function is defined as: 
\[
\Tilde{s}_v^\infty = \begin{cases}
{\sf activate}\left(\bm{W_h h_v^k}\Vert \bm{W_d d_m}\right) & v\not\in C_m^T\\
1 & v\in C_m^T
\end{cases}
\]
where function {\sf activate} is implemented as a 3-layer neural network in this paper 
and  $\bm{W_h}$ and $\bm{W_\delta}$ are two parameters matrix to be learned. 
We add this function as an additional layer after the last layer of the GCN and GAT. 

\medskip\noindent\textbf{SE-CGNN.} Recall that CoupledGNN uses the function 
{\sf InfluGate} to simulate the process of a user to be activated by their neighbours. 
The influence vector, e.g., $r_u$ of user $u$, contains user $u$'s posted message 
and the messages from $u$'s neighbourhood. Therefore, it can be considered as a
summary of the information perceived by a user $v$ from $u$ if $v$ follows $u$ in Twitter.
Based on this intuition, we extend CoupledGNN by reformulating the function 
{\sf InfluGate(*)} to capture the the info-exposure spillover effect:
\[
{\sf influGate}\left(\bm{r_u^\ell}, \bm{r^\ell_v}\right) =
\bm{\beta^\ell}\left[\bm{W^\ell r_u^\ell}\Vert \bm{W^\ell r_v^\ell}\Vert
\bm{W_d d_m}\right].
\]
\subsection{Objective function}
We use the same objective function of~\cite{cao2020popularity} which is the mean
relative square error (MRSE) and defined as the follows:
\[
L_{\it MRSE} = \frac{1}{M}\sum_{n=1}^M \left(\frac{\Tilde{n}_\infty^m-n_\infty^m}{n_\infty^m}\right)^2.
\]
This loss function is regularised to avoid over-fitting and accelerate the convergence
speed, i.e., $L= L_{\it MRSE} + L_{\it Reg}$ where 
$L_{\it Reg} = \theta\sum_{p\in\mathcal{P}}\Vert p\Vert_2+ \lambda L_{\it user}$.
Note that $\mathcal{P}$ denotes the set of parameters and $L_{\it user}$ is the
cross-entropy 
$\frac{1}{M}\sum_{n=1}^{M}\frac{1}{\mid\mathcal{V}\mid}\left(
s_v^\infty\log s_v^k + (1-s_v^\infty)\log(1-s_v^k)
\right)$ where $s_v^\infty$ is the final activation status of $v$.

\section{Experimental Evaluation}
\label{sec:experiment}

\subsection{Evaluation metrics}
We adopts the metrics used in~\cite{cao2020popularity} to evaluate and compare the 
performance of our extended models and the bench-markings used in our experiments. 
Specifically, in addition to the mean relative square error (MRSE) introduced in 
the previous section, we also use mean absolute percentage error (MAPE) and 
wrong percentage error (WroPerc). MAPE measures the average deviation between 
the predicted popularity and the true one while 
WroPerc measures the percentage of cascades that are incorrectly predicted with 
a given error tolerance $\epsilon$.
formally, they can be formally defined as:
\begin{displaymath}
\begin{array}{ll}
 {\it MAPE}&= \frac{1}{M}\sum_{m\in \mathcal{M}_C} 
 \frac{\mid\Tilde{n}_\infty^m- n_\infty^m\mid}{n_\infty^m},\\
  {\it WroPerc}& = \frac{1}{M}\sum_{m\in \mathcal{M}_C}\mathbb{I}\left[ \frac{\mid\Tilde{n}_\infty^m- n_\infty^m\mid}{n_\infty^m} \ge \epsilon\right].
\end{array}
\end{displaymath}
Note that $\mathbb{I}(*)$ is an indication function which outputs $1$ when the input proposition 
is true or 0 otherwise, and the threshold $\epsilon$ is set as $0.5$ in our experiments. 

\subsection{Baseline Methods}

In addition to CoupledGNN, we use the following models as baselines.

\noindent\textbf{Feature-based method}. This is a linear regression model with L2 
regularisation with features. For better comparison, we adopt the same features used in the 
past studies~\cite{cao2020popularity,li2017deepcas}. 

\noindent\textbf{SEISMIC}~\cite{zhao2015seismic}. 
SEISMIC uses the Hawkes self-activation point process to estimate or approximate the 
impact of cascading effect by their average number of followers.

\noindent\textbf{DeepCas}~\cite{li2017deepcas}. 
DeepCas is an end-to-end deep learning method for information cascades prediction. 
It utilises the structure of the cascade graphs and node identities for prediction. 
An attention mechanism is designed to assemble a cascade graph representation from a set 
of random walk paths. 

\noindent\textbf{GCN and GAT}. We construct these two models from our 
SE-GCN and SE-GAT models by removing the representation vectors of messages.
In other words, these two models only rely on network structure to predict 
the size of final cascades.

\subsection{Implementation details}
\label{subsec:Implementation}
As the output of the RoBERTa for a sentence is a  high-dimensional and sparse vector. 
we apply linear transformation to map its output to a relatively low-dimensional space. 
The dimension of the final text embedding used is set at $128$. 
For all models including baselines, we tune their hyper-parameters to ensure a good performance
on validation sets. The L2-coefficients are chosen from $\{0.5,0.1,0.05,\cdots,10^{-8}\}$. 
For all neural network models, the learning rate is chosen from $\{0.1,0.05,\cdots,10^{-5}\}$, 
the coefficient in loss function is set to be 0.5, and 
the mini-batch size is chosen from $\{15,10,5\}$. 
The number of GNN layers $k$ is selected from $\{5,4,3,2\}$. 
As for DeepCas, the number of walk sequences with walk length are set as $100$ and $8$, respectively. 
For SEISMIC, we follow the parameters from the original study, i.e. setting the constant 
period as $5$ minutes and power-law decay parameters $\theta$ as  $0.242$.  
Considering the diffusion time of the messages in our collected data, 
we set the observation time window $T$ as 3 hours and construct a set of observed cascades by removing users in our cascades that were 
activated after the first 3 hours. 

In order to comprehensively evaluate the effectiveness of info-exposure spillover effect in predicting 
the message popularity, in addition to the cascades of COVID-19 preventive measure messages 
$\mathcal{C}_\PM$, we also 
apply all the models on another two sets of cascades. One is the set of all COVID-19 related 
cascades $\mathcal{C}$. The other is the set of COVID-19 related cascades that are not
related to preventive measures, i.e., 
$\overline{\mathcal{C}}_{\PM}=\mathcal{C}/\mathcal{C}_{\PM}$, the complement 
of $\mathcal{C}_{\PM}$ in $\mathcal{C}$.

\subsection{Experiment results}
\begin{table*}[tp]
\caption{Prediction results}
\label{table:results}
\centering
\begin{tabular}{|l|r|r|r|r|r|r|r|r|r|}
\hline
\rule{0pt}{9pt} \multirow{2}{*}{\textbf{Models}}  & \multicolumn{3}{c|}{\textbf{$\mathcal{C}$}} & 
 \multicolumn{3}{c|}{\textbf{${\mathcal{C}_{\it PM}}$}}  & 
 \multicolumn{3}{c|}{\textbf{$\overline{\mathcal{C}}_{\it PM}$}}   \\ \cline{2-10} 
 &
 \rule{0pt}{9pt} \textbf{MRSE} & \textbf{MAPE} &\textbf {WroPerc} & \textbf{MRSE} &
  \textbf{MAPE} & \textbf{WroPerc} & \textbf{MRSE} & \textbf{MAPE} & \textbf{WroPerc} \\ \hline\hline
Feature-based & 0.3611 & 0.4018 & 41.31\% & 0.4403 & 0.4049 & 46.08\% & 0.3704 & 0.4151 & 41.56\% \\ \hline
SEISMIC       & 0.5580 & 0.5104 & 56.35\% & 0.5899 & 0.5265 & 55.88\% & 0.5419 & 0.5083 & 56.14\% \\ \hline
DeepCas       & 0.2837 & 0.3959 & 37.71\% & 0.2847 & 0.3724 & 38.67\% & 0.2872 & 0.4010 & 37.31\% \\ \hline
GCN           & 0.3144 & 0.4217 & 38.88\% & 0.3179 & 0.4238 & 41.76\% & 0.3110 & 0.4200 & 38.69\% \\ \hline
SE-GCN        & 0.2826 & 0.4056 & 36.76\% & 0.2702 & 0.3961 & 35.44\% & 0.2899 & 0.4109 & 36.65\%   \\ \hline
GAT           & 0.3072 & 0.4211 & 39.19\% & 0.3014 & 0.4268 & 40.01\% & 0.3101 & 0.438  & 39.85\% \\ \hline
SE-GAT        & 0.2862 & 0.4124 & 37.58\% & 0.2721 & 0.4001 & 35.31\% & 0.2903 & 0.4175 & 38.64\% \\ \hline
CoupledGNN    & 0.2678 & 0.3861 & 35.19\% & 0.2769 & 0.3920 & 34.44\% & 0.2601 & 0.3812 & 34.70\% \\ \hline
SE-CGNN &
  \textbf{0.2214} &
  \textbf{0.3410} &
  \textbf{31.17\%} &
  \textbf{0.2087} &
  \textbf{0.3001} &
  \textbf{28.13\%} &
  \textbf{0.2261} &
  \textbf{0.3508} &
  \textbf{31.22\%} \\ \hline
\end{tabular}
\end{table*}

We list the performance of all the above mentioned models in Table~\ref{table:results} 
in the form of the three selected metrics. In general, we can see two obvious differences
when the info-exposure spillover effect is introduced in cascade prediction. 

First, compared to the original models, our extended models significantly improve their performance
not only for the preventive measure messages, but also for all the three types of messages.
The most significant improvement occurs to SE-CGNN and reaches $23\%$ in the WroPerc measurement for 
the preventive measure messages and over $10\%$ for the messages unrelated to preventive measures. 
This is due to the fact that CoupledGNN simulates the cascading effects iteratively and this 
allows for applying the info-exposure spillover effect on activating individual users in a finer granularity. 
From the above analysis, we can conclude that the use of info-exposure spillover effects can 
effectively improve the performance of existing cascade prediction models. It should be 
integrated into future models by design. 

Second, we can observe that the extended models can more accurately predict the popularity of 
COVID-19 preventive measure messages than the other messages, which is the opposite for 
the baseline models. 
From the baseline models, we see that it is more difficult for accurately predict 
the final cascade size of COVID-19 preventive measure messages. Their performance on $\mathcal{C}$
and $\overline{\mathcal{C}}_{\it PM}$ are almost the same but becomes worse on $\mathcal{C}_{\it PM}$.
The feature-based model has the worst performance which decreases by over $11\%$ compared to that in 
predicting the size of the other two sets of cascades. However, when the identified info-exposure 
effect is used in our extended models, the popularity of preventive measure messages can be 
predicted with a much better accuracy. SE-CGNN can improve the performance in the set $\mathcal{C}_{\it PM}$ 
by about $10\%$ better for preventive measurement messages than those unrelated to preventive messages.
This observation validated empirically that the exposure to information related
to the COVID-19 pandemic has a strong spillover effect on retweeting message related to COVID-19 
messages about how to prevent the transmission of the virus.

\section{Conclusion \& Discussion}
\label{sec:Discussion}

In this paper, we concentrated on the problem of cascade prediction for COVID-19 information about 
preventive measures on online social media platforms. Compared to previous works, we took into 
account the phenomenon that the exposure to COVID-19 information will influence the behaviour of 
users to participate in the diffusion of information related to preventive measures, which we call 
\emph{info-exposure spillover effect} in this paper. 
With a dataset we collected from Twitter, we successfully validated its existence. 
We then applied the identified
spillover effects in predicting the popularity of preventive measure messages. 
Specifically, we built three new models by making use of the recent advances of graph representation 
techniques, i.e., graph neural networks. 
With extensive experiments, we showed that our new models outperform baselines not only for 
preventive measure messages but for all the COVID-19 related messages. 
This illustrates that the introduction of info-exposure spillover effect can effectively improve the 
performance of cascade prediction. 

There are still several limitations in our research. When representing users' historical 
textual posts, we took the mean of their representation vectors. 
This may remove certain useful information hidden in users' past messages.
Moreover, we ignore the significance variance caused by the post time of messages. 
It has been studied that recent messages may have larger influence. 
This can be solved by introducing recurrent networks such as LSTM or the Hawkes process. 
Second, our cascade prediction models are extended from existing GNN models and 
focus on preventive measure messages. 
It will be an interesting future work to design a new general end-to-end GNN model which 
can capture diffusion patterns shared by COVID-19 related messages.  

\medskip
\noindent{\bf Acknowledgements.}
This work was partially supported by Luxembourg’s Fonds National de la Recherche, via grant {\sf COVID-19/2020-1/14700602} (PandemicGR) and grant PRIDE17/12252781/DRIVEN. 

\bibliographystyle{IEEEtran}
\bibliography{ref}

\end{document}